\documentclass[a4paper]{jpconf}

\newcommand{\be}{\begin{equation}}
\newcommand{\ee}{\end{equation}}
\newcommand{\bra}{\langle}
\newcommand{\ket}{\rangle}
\newcommand{\bea}{\begin{eqnarray}}
\newcommand{\eea}{\end{eqnarray}}
\newcommand{\dis}{\displaystyle}

\usepackage{graphicx}
\begin{document}
\title{ Large-Scale Simulation of Multi-Asset Ising Financial Markets}

\author{Tetsuya Takaishi${}^1$}

\address{$~^1$Hiroshima University of Economics, Hiroshima 731-0192, Japan \\
}

\ead{tt-taka@hue.ac.jp}

\begin{abstract}
We perform a large-scale simulation of an Ising-based financial market model 
that includes 300 asset time series.
The financial system simulated by the model shows a fat-tailed return distribution and volatility clustering
and exhibits unstable periods indicated by the volatility index measured as the average of absolute-returns.  
Moreover, we determine that the cumulative risk fraction, which measures the system risk, changes 
at high volatility periods.
We also calculate the inverse participation ratio (IPR) and its higher-power version, IPR6,
from the absolute-return cross-correlation matrix. Finally, we show that the IPR and IPR6 also change at high volatility periods.
\end{abstract}

\section{Introduction}
Statistical properties of asset prices have been intensively studied, and
some pronounced properties such as the fat-tailed return distribution and 
volatility clustering have been investigated.
These properties are now classified as ``stylized facts" , e.g., see \cite{CONT}.
Possible dynamics for asset returns  
have been given by Clark \cite{Clark}.
He suggested that return dynamics follow a Gaussian random process with time-varying volatility,
also known as the mixture-of-distribution hypothesis (MDH).
Let $r_t$ be the return at time $t$; 
the return is described by $r_t=\sigma_t \epsilon_t$,
where $\sigma_t^2$ is the time-varying volatility and $\epsilon_t$ is a standard normal value $\sim N(0,1)$.
Under the MDH, the volatility varies according to the rate of information arrival to the market.
Since it is difficult to estimate the rate of information arrival in real financial markets,
volume was used as a proxy \cite{Clark}. 

Determining whether the return dynamics follow a Gaussian random process with time-varying volatility 
can be verified by examining the returns standardized by $\sigma_t$.
The standardized returns are given by $\bar{r}_t =r_t/\sigma_t$.
If the return is described by $r_t=\sigma_t \epsilon_t$, then
the standardized returns will be $\epsilon_t$, and 
the normality for  $\bar{r}_t$ should be observed, e.g., the variance is equal to one and the kurtosis is equal to three.
A drawback of this verification is that the volatility cannot be directly obtained in real financial markets.
Recent availability of high-frequency intraday returns enables us to construct realized volatility (RV) \cite{RV,RV2,RV3},
which converges to the integrated volatility at the infinite sampling  frequency. 
Empirical studies using the RV claim that the return dynamics are approximately consistent with 
a Gaussian random process with time-varying volatility \cite{RV2,RV3,RV4,RV6,Takaishi,T-W}.

The RV has different properties from other volatility measures such as absolute volatility, which is often used by econophysists.
The clustering and memory effects of the RV have been investigated, and the RV has also been compared with the absolute volatility \cite{PLOSONE}. The distribution of the RV has also been studied, and it has been argued that the RV distribution
is well described by an inverse gamma distribution \cite{Super}. 

To better understand the price dynamics observed in real financial markets,
Bornholdt proposed a simple and minimalistic Ising-based spin model that includes only two interactions
that conflict with each other \cite{Born}.
One of the interactions corresponds to the majority effect that agents imitate their neighbors; the other is the effect that 
agents tend to join minority groups.
The model with these conflicting interactions shows non-equilibrium dynamics in return time series 
and successfully exhibits major stylized facts such as fat-tailed return distributions 
and volatility clustering \cite{Born,Born2,POTTS,Born3}.
The return dynamics with time-varying volatility can be verified by calculating $R_t/\sigma_t$; moreover, it has been shown
that $R_t/\sigma_t$ recovers the same standard normality as real financial markets \cite{RVspin1,RVspin2}.

The original model by Bornholdt can only simulate the dynamics of one asset price.
Real financial markets include many stocks correlated with each other. 
Measuring correlations between asset returns is especially important to the study of financial market stability,
and quit a few empirical studies have been conducted that unveil properties of correlations among asset returns
in real financial markets, e.g., \cite{CC0,CCC0,CCC1,CC1,CC2,CC3,CC4}.

To simulate multiple stock time series, the Bornholdt model was extended in \cite{Multiple1,Multiple2,Multiple3},
and it is shown that the extended model exhibits major stylized facts 
similar to the original Bornholdt model.
In this study, to further investigate the properties of the extended model,
we construct a large-scale simulation including 300 assets
and study dynamical properties of correlations and the instability of the system
simulated by the extended model. 

\section{Extended Ising-based Financial Model}
Originally, the Ising-based model was introduced by Bornholdt \cite{Born} 
and then extended to simulate multiple time series in \cite{Multiple1,Multiple2,Multiple3}. 
We consider a financial market where $N$ stocks are traded and
assume that spin agents are located on sites of an $L \times L$ square lattice as in \cite{Born}.
The number of total sites on a lattice is $P=L\times L$.
Each site of the lattice has a spin agent $s_i$ that takes the value $+1$ or $-1$,  where $i$ stands for the $i$-th agent;
$s_i=+1$ $(-1)$ denotes that the agent is assigned the ``Buy" (``Sell") state.

The agents flip their spin states probabilistically according to a local field.
The local field of the $i$-th spin $h_i^{(k)}(t)$ at time $t$ for the $k$-th stock is defined by
\be
h_i^{(k)}(t)=\sum_{\bra i,j\ket}J s_j^{(k)}(t) -\alpha^{(k)} s_i^{(k)}(t)|M^{(k)}(t)|  + \sum_{j=1}^N \gamma_{jk}M^{(j)}(t),
\label{eq:h}
\ee
where $\bra i,j\ket$ stands for a summation over the nearest neighbor pairs and
$J$ is the nearest neighbor coupling. In this study, we set $J=1$, and
$M^{(k)}(t)$ is the magnetization
defined by $\dis M^{(k)}(t)=\frac1P\sum_{l=1}^P s_l^{(k)}(t)$.

The first term on the right-hand side of (\ref{eq:h}) with $J>0$ 
introduces the ferromagnetic effect similar to the original Ising model 
that tends to align nearest neighbor spins with the same sign.
The second term introduces the effect that promotes a spin-flip, which corresponds to the minority effect.
In this study, we assume that all $\alpha^{(k)}$ have the same value; that is, $\alpha^{(k)} \equiv \alpha$.
The third term that is not present in the Bornholdt model describes the interaction with other stocks and
introduces the effect of imitating the states of other stocks.
We assume that the magnitude of the interaction is given by the interaction parameters that form a matrix $\gamma_{lm}$
that has zero diagonal elements, i.e., $\gamma_{ll}=0$.
We update the spin states according to the following probability $p$:
\bea
\label{eq:Prob}
s_i^{(k)}(t+1) & =   +1  & \hspace{5mm}  p=1/(1+\exp(-2\beta h_i^{(k)}(t))), \\ \nonumber
s_i^{(k)}(t+1) & =  -1   & \hspace{5mm} 1- p,
\eea
where $\beta$ is a parameter that corresponds to the inverse temperature in the original Ising model.

\section{Simulation}
In this study, we consider a financial system that trades 300 stocks.
Each stock is traded on a 100$\times$100 square lattice with a periodic boundary condition.
The simulation parameters are set to $(\beta,\alpha)=(2.3,60)$.
Since the situation where all elements $\gamma$ are non-zero is unrealistic, 
we assume that ten percent of the off-diagonal elements in $\gamma$ are non-zero and 
set to non-zero values drawn from Gaussian random numbers with average 0.05 and variance 0.01.
The remaining elements are set to zero.

Spins are updated according to (\ref{eq:Prob}) in random order.
We begin the simulation on a lattice with ordered spins and then discard the first $5\times10^3$ sweeps
as thermalization.
Then, we collect data from $3\times10^4$ updates for analysis. 

Following \cite{Born2}, the return of the $k$-th stock is defined by the difference of the magnetization,
\be
R_k(t)=(M^{(k)}(t+1)-M^{(k)}(t))/2,
\label{return}
\ee
where $t$ is incremented in units of one update.
In Fig. 1, we show a representative return time series that exhibits volatility clustering,
which is one of the stylized facts often seen in real financial markets \cite{Mantegna,Stanley2}. 
Figure 2 shows the autocorrelation functions of return, absolute-return, and squared-return.
Notice that the autocorrelation of return quickly disappears with time.
On the other hand, the autocorrelations of absolute-return and squared-return remain finite at large time
and slowly converge to zero within the noise level.
The integrated autocorrelation times are estimated to be $223\pm112$ for the absolute-return and $93\pm33$ for the squared-return.
These properties of autocorrelation are also observed in real financial markets \cite{CONT}.

\begin{figure}
\centering
\includegraphics[height=6.5cm,width=10cm]{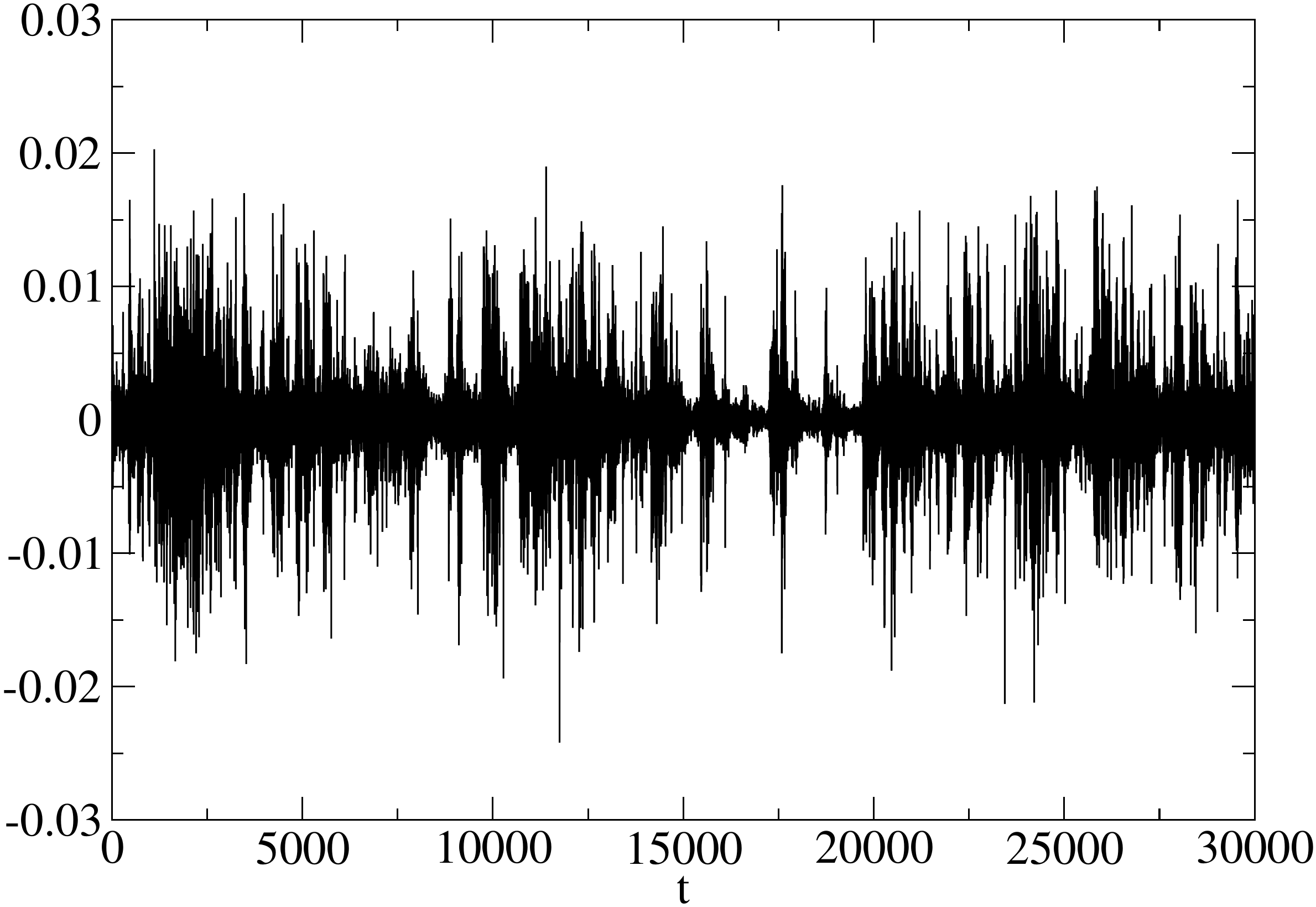}
\caption{
Representative return time series $R_k(t)$ simulated from the extended model, where $t$ is incremented in units of one update.
}
\label{fig:History}
\end{figure}

\begin{figure}
\vspace{5mm}
\centering
\includegraphics[height=6.5cm,width=10cm]{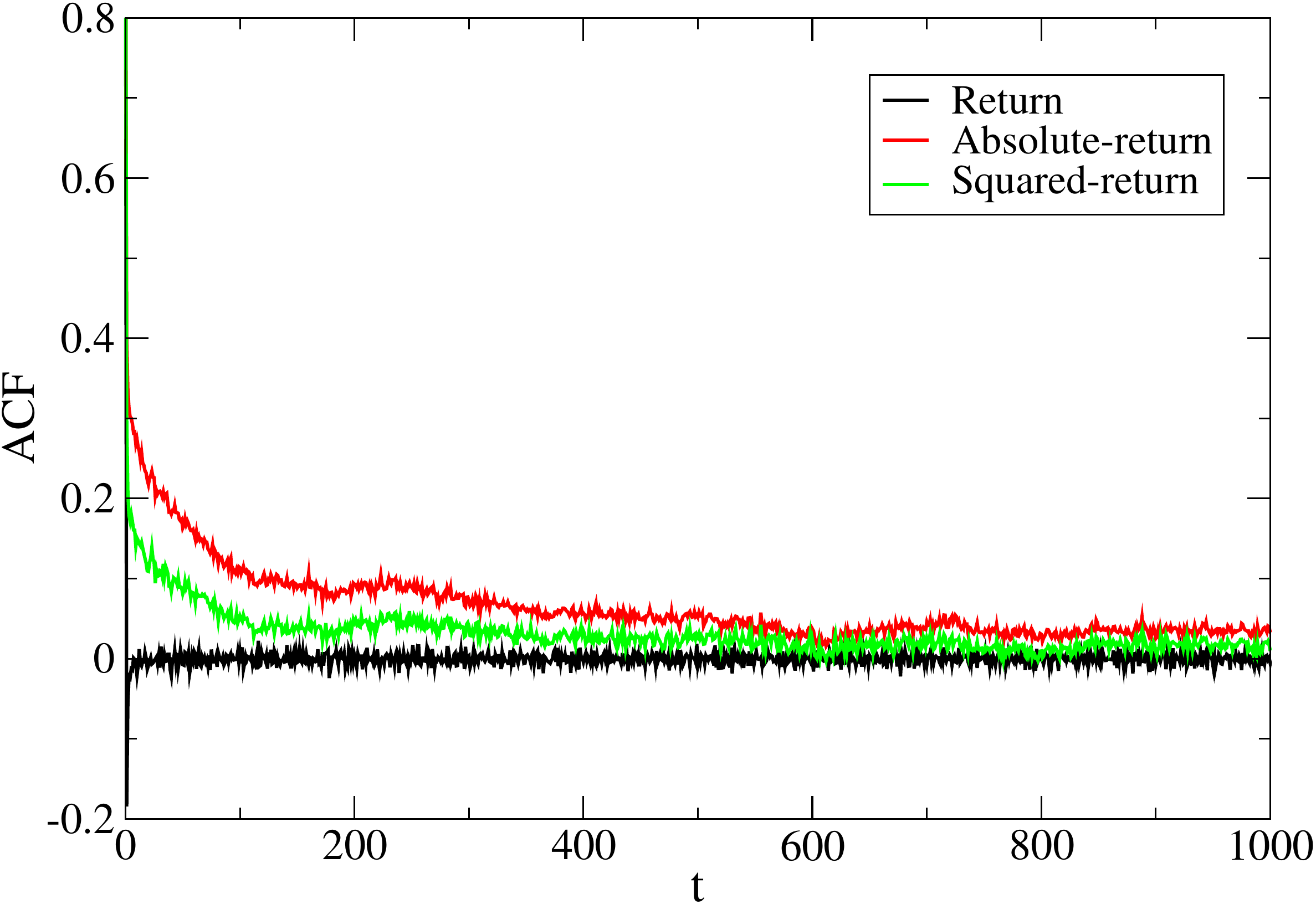}
\caption{
Autocorrelation functions of return, absolute-return, and squared-return as a function of $t$, where $t$ is incremented in units of one update.
}
\label{fig:History}
\end{figure}

Figure 3 shows the return distribution constructed from
the normalized returns. Each return time series is normalized according to $\bar{R}_k(t)=(R_k(t)-A_k)/\sigma_k$,
where $A_k$ is the average value of $R_k(t)$ and $\sigma_k$ is the standard deviation of $R_k(t)$.
Notice that the return distribution is fat-tailed.
In real financial markets, this fat-tailed nature of return distributions
has been investigated in literature such as in \cite{ Stanley2, Lux, Stanley1}.

\begin{figure}
\vspace{5mm}
\centering
\includegraphics[height=6.5cm]{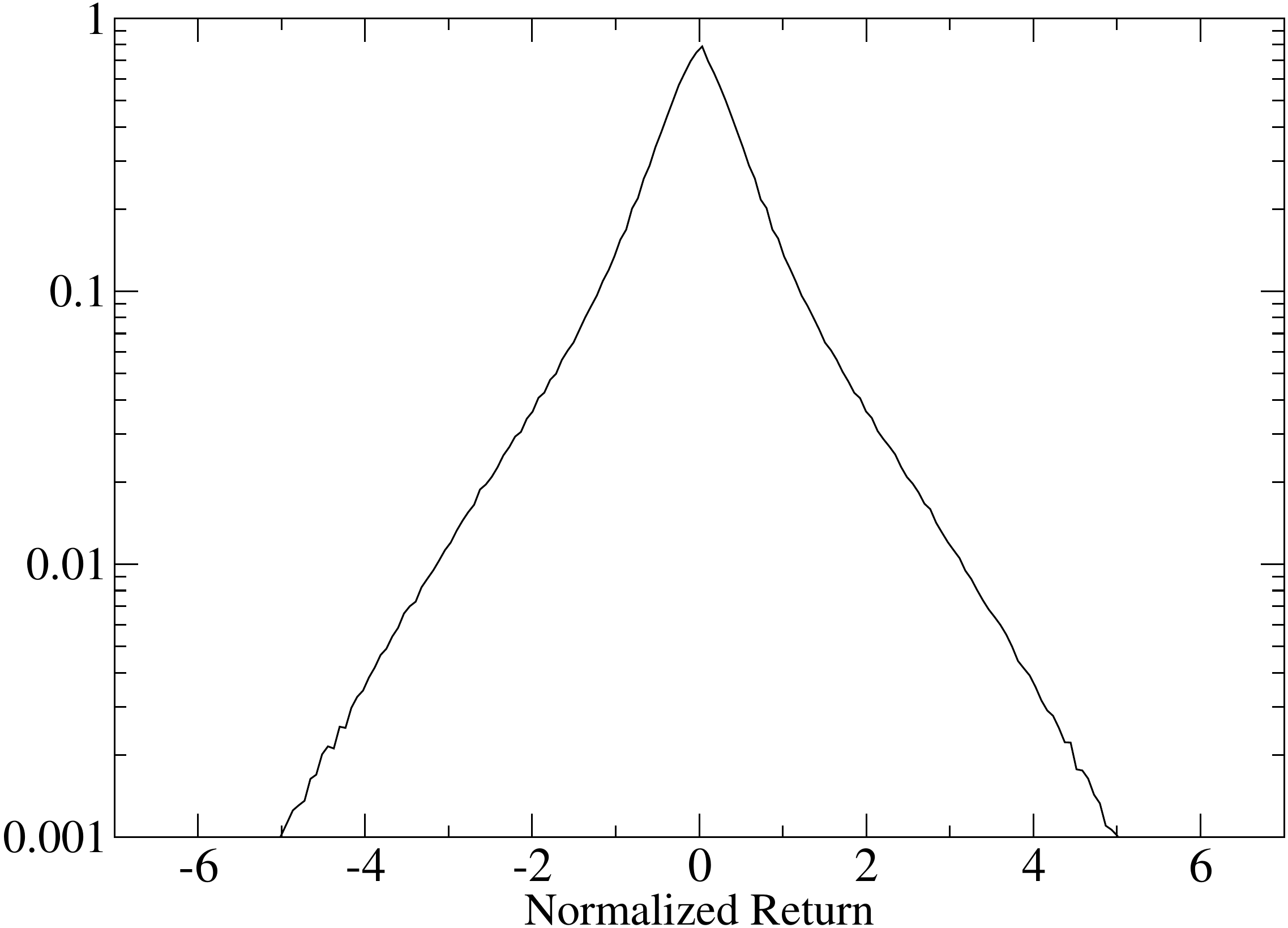}
\caption{
Normalized return distribution.
}
\label{fig:History}
\end{figure}

To dynamically quantify the volatility level of the financial system simulated according to the model,
we define a volatility index by the average of $|R_k(t)|$:
$ \dis I(t)=\frac1N \sum_{k=1}^N |R_k(t)|$.
We determined  that the volatility index $I(t)$ shown in Fig. 4 exhibits 
high and low periods.
We expect the system to experience unstable periods  
when the volatility index is high.

\begin{figure}
\centering
\includegraphics[height=6.5cm,width=12cm]{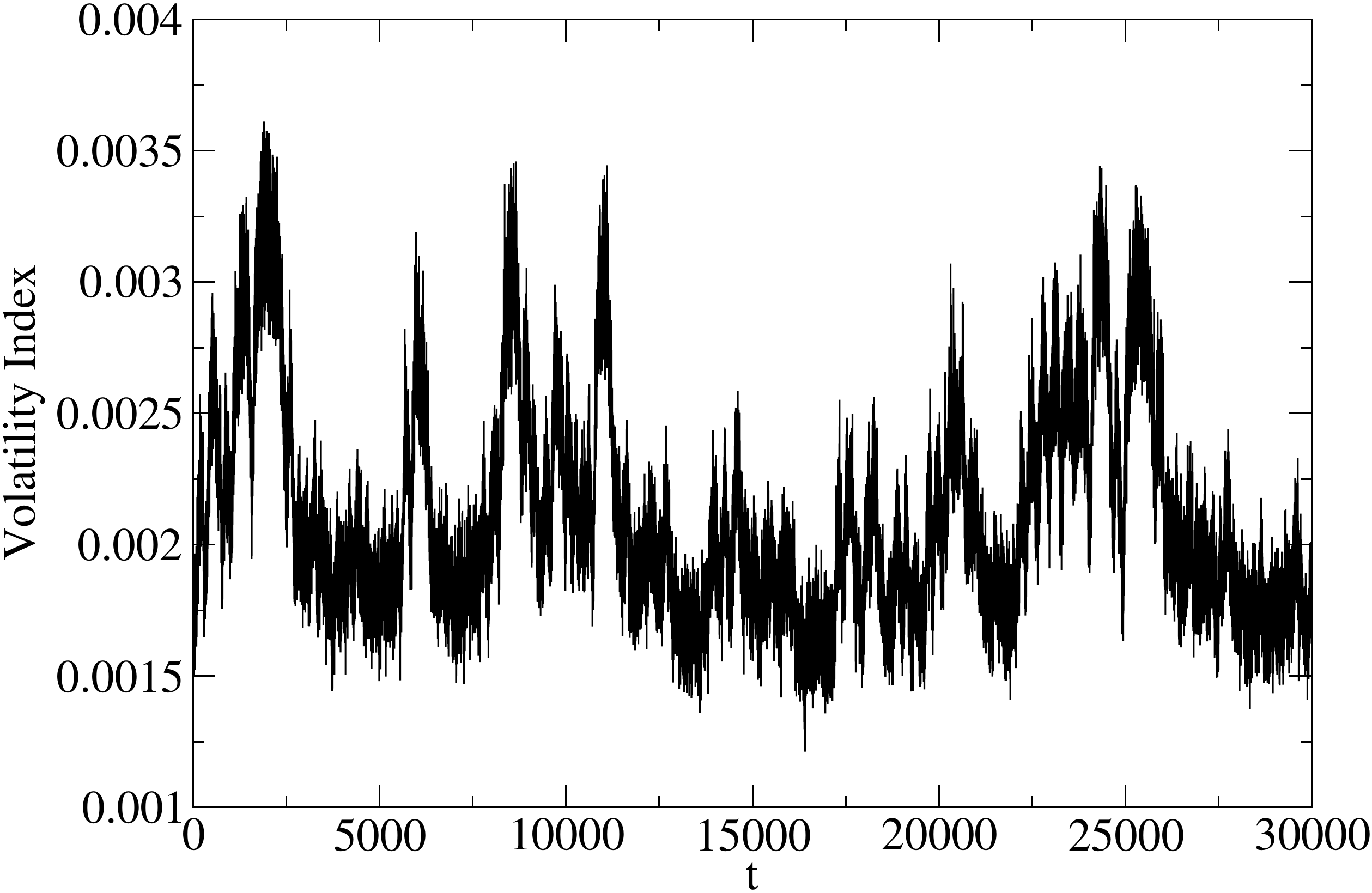}
\caption{
Volatility index $I(t)$.
}
\label{fig:History}
\end{figure}

\section{Cross-correlation Matrix}
To investigate the instability of the system, we calculate the cross-correlation matrix.
Let $R_k(t)$ be a return for stock $k$ $(k=1,...,N)$ at time $t$ $(t=1,...,T)$ 
defined by (\ref{return}), where $N=300$ and $T=30000$.
We also define
the normalized return $m_k(t)$ by
\be
m_k(t)=\frac{R_k(t)-\bra R_k \ket}{\sigma_k},
\ee
where $\bra ... \ket$ indicates the time series average and
$\sigma_k$ is the standard deviation of $R_k(t)$.
Using the normalized return $m_k(t)$,
an equal-time cross-correlation matrix $C(t)$ is 
defined by
\be
C_{kj}(t) =
\frac1M\sum_{i=0}^{M-1}m_k(t-i) m_j(t-i),
\label{eq:CC}
\ee
where  $M=400$, i.e.,  the average in (\ref{eq:CC}) is taken over a period of 400.
In this study, we also calculate the equal-time cross-correlation matrix from the absolute-returns.

\section{Cumulative Risk Fraction}
In order to investigate the systemic risk in the financial system,
Billio {\it et al.} \cite{SR1} suggested using principal component analysis (PCA)
and introduced the cumulative risk fraction (CRF) as a risk measure.
The CRF has also been studied in \cite{SR2,SR3,SR4,SR5}.
Here, we use the CRF to study the dynamical evolution of instability of the system.

The CRF is calculated as follows.
First, we compute the eigenvalues of the cross-correlation matrix, denoted by $\lambda_1,\lambda_2,\dots,\lambda_N$, such that
$\lambda_1>\lambda_2>\dots>\lambda_N$.
Then, the CRF is defined by \cite{SR1}:
\be
CRF_m= \frac{\omega_m}{\Omega},
\ee
where
$\Omega$ is the total variance of the system given by
$\Omega= \sum_{j=1}^N \lambda_j$ and
$\omega_m$ is the risk associated with the first $m$ principal components,
given by $\omega_m= \sum_{j=1}^m \lambda_j$.
Here, note that for cross-correlation matrices, $\Omega=1$.
The CRF quantifies the portion of the system variance explained by the first $m$ principal components
over the total variance \cite{SR2}.
In periods of financial crisis, many stocks are highly correlated with each other 
and their prices easily move together.
As a result, the volatility of stocks also increases,
and the CRF is expected to increase in such financial crisis periods.

Figure 5(a) shows $CRF_1$ to $CRF_5$ calculated from 
the return cross-correlation matrix.
Compared to the volatility index $I(t)$ in Fig. 4, notice that 
the CRF increases in high volatility periods.
Figure 5(b) is the same as  Fig. 5(a) but for the absolute-return cross-correlation matrix.
Contrary to the CRF from the return cross-correlation matrix,
the CRF from the absolute-return cross-correlation matrix decreases in high volatility periods.

\begin{figure}
\centering
\includegraphics[height=7.5cm]{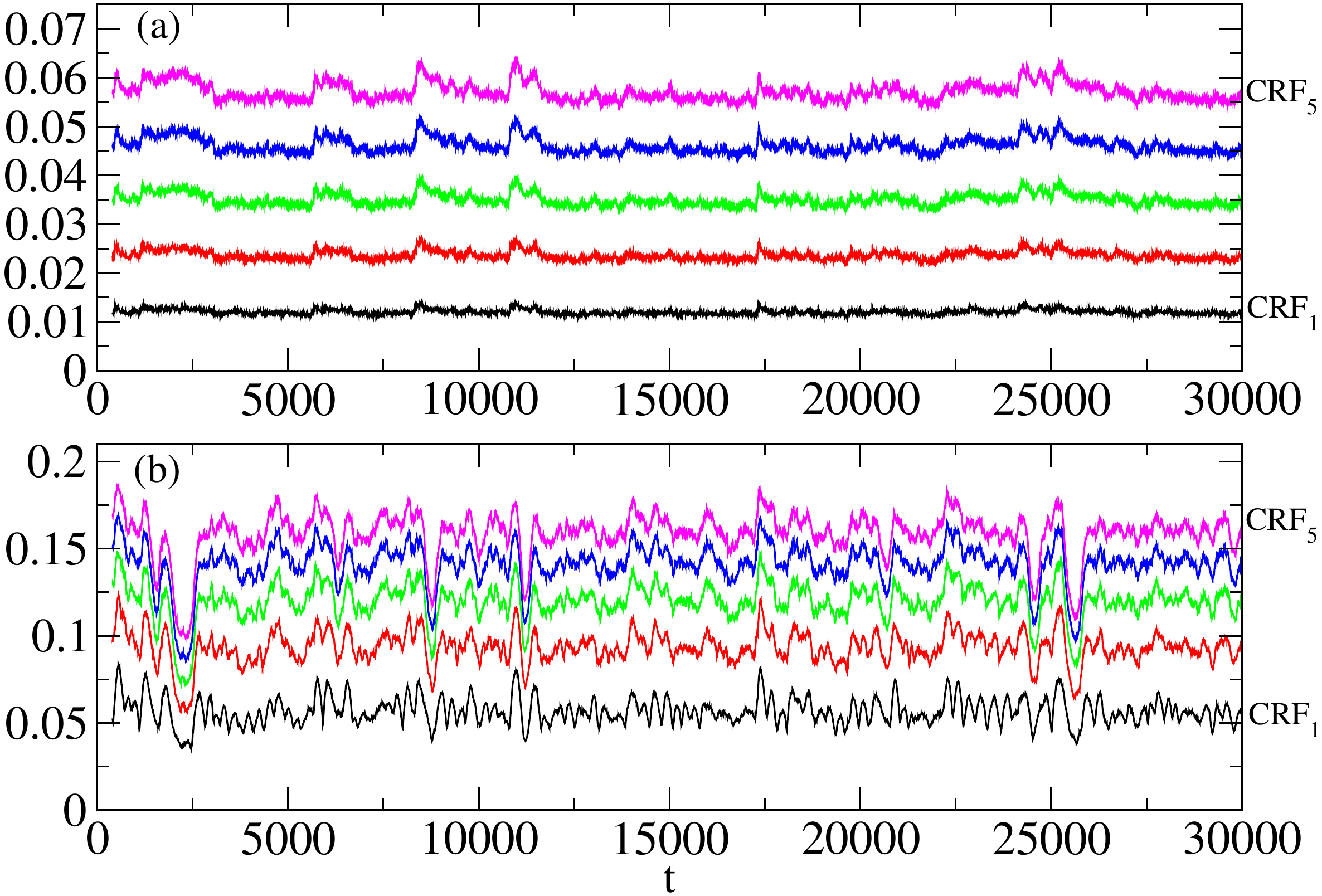}
\caption{
Cumulative risk fraction from the return (a) and absolute-return (b).
}
\label{fig:History}
\end{figure}

\section{Random Matrix Theory}

To further investigate the dynamical instability of the system, we utilize 
results from random matrix theory (RMT).
Let $y_i(t)$ be an independent, identically distributed random variable with $i=1,...,N$ at time $t=1,...,T$.
Then, we define the normalized variable
\be
w_i(t)=\frac{y_i(t)-\bra y_i \ket}{\sigma_{y_i}},
\ee
where $\sigma_{y_i}$ is the standard deviation of $y_i$.
The equal time cross-correlation between variables $y_i(t)$
is given by $ W_{ij}=\bra w_i w_j \ket$.
The matrix $W$ is called the Wishart matrix.
For $N\rightarrow \infty$ and $T\rightarrow \infty$ with $Q=T/N>1$, 
the eigenvalue distribution of the matrix $W$ is known in RMT \cite{W1,W2} as
\be
\rho(\lambda)=\frac{Q}{2\pi}\frac{\sqrt{(\lambda_+ -\lambda)(\lambda-\lambda_- )}}{\lambda}, \hspace{3mm}
\lambda_\pm =1+\frac1Q\pm 2\sqrt{\frac1Q}.
\ee
For $T=400$ and $N=300$, we obtain $\lambda_+=3.482$ and $\lambda_-=0.01795$.

The inverse partition ratio (IPR) that characterizes the eigenvectors is defined by 
$ \dis IPR(l) =\sum_{j=1,}^N (v_l^j)^4$,
where $v_l^j$ is the $j$-th component of the eigenvector for the $l$-th eigenvalue.
RMT predicts that  the eigenvector components are de-localized and distributed as a Gaussian distribution
$\dis \sim \sqrt{\frac{N}{2\pi}}\exp\left(-\frac{N}{2}(v_l^j)^2\right)$.
In such a case, the expectation of the IPR is given by $3/N$. On the other hand, when the eigenvector components are localized, for example, if only one component has a non-zero value,
the expectation of the IPR is 1.

The IPR can be extended to its higher-power version, IPR6.
In this study, we also focus on IPR6 defined by
$ \dis IPR6(l) =\sum_{j=1,}^N (v_l^j)^6$.
The expectation of IPR6 in RMT is given by $15/N^2$.

\begin{figure}
\centering
\includegraphics[height=7.5cm]{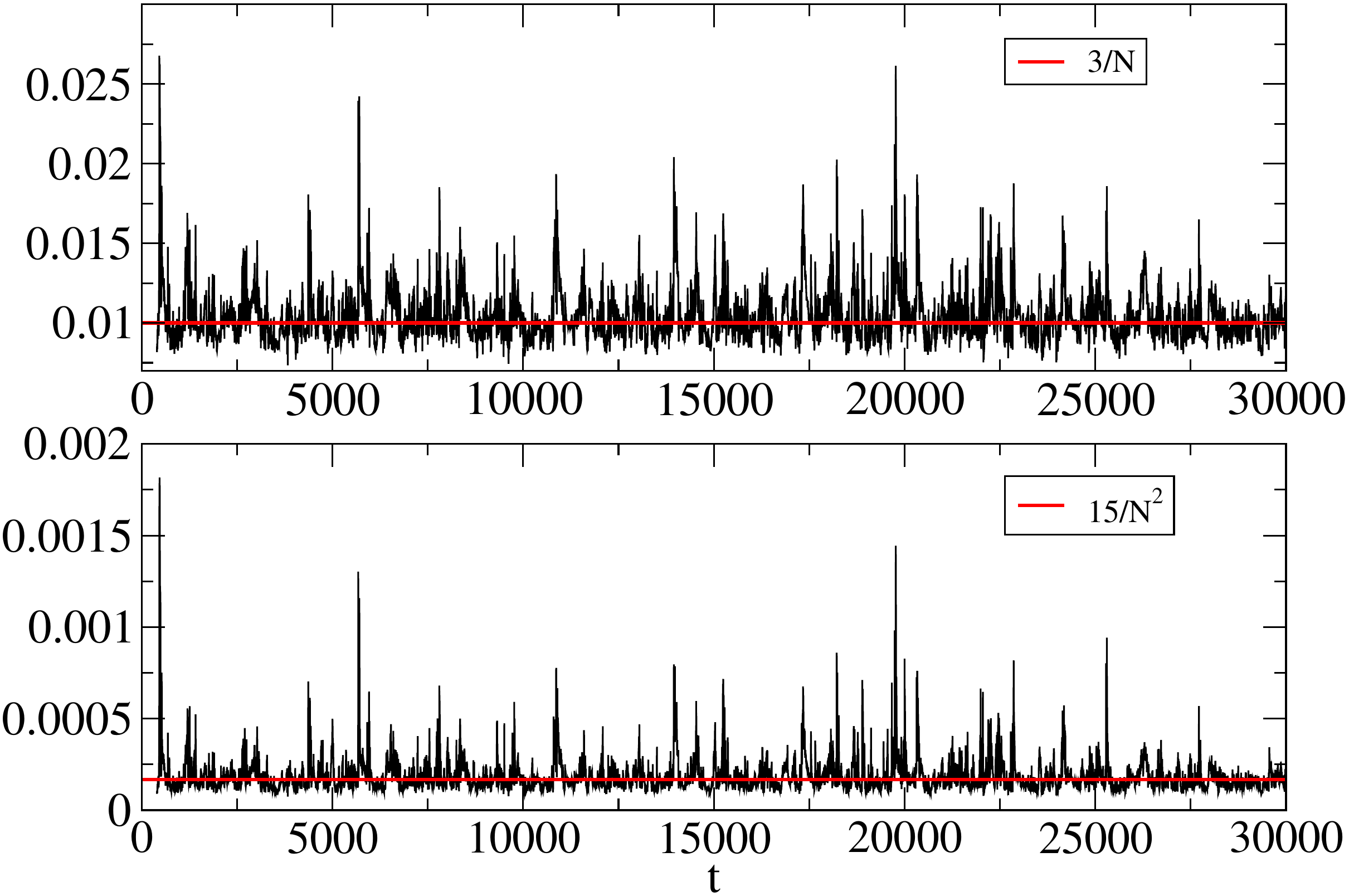}
\caption{
$IPR(1)$ and $IPR6(1)$ from return eigenvectors.
}
\label{fig:History}
\end{figure}

\begin{figure}
\vspace{5mm}
\centering
\includegraphics[height=7.5cm]{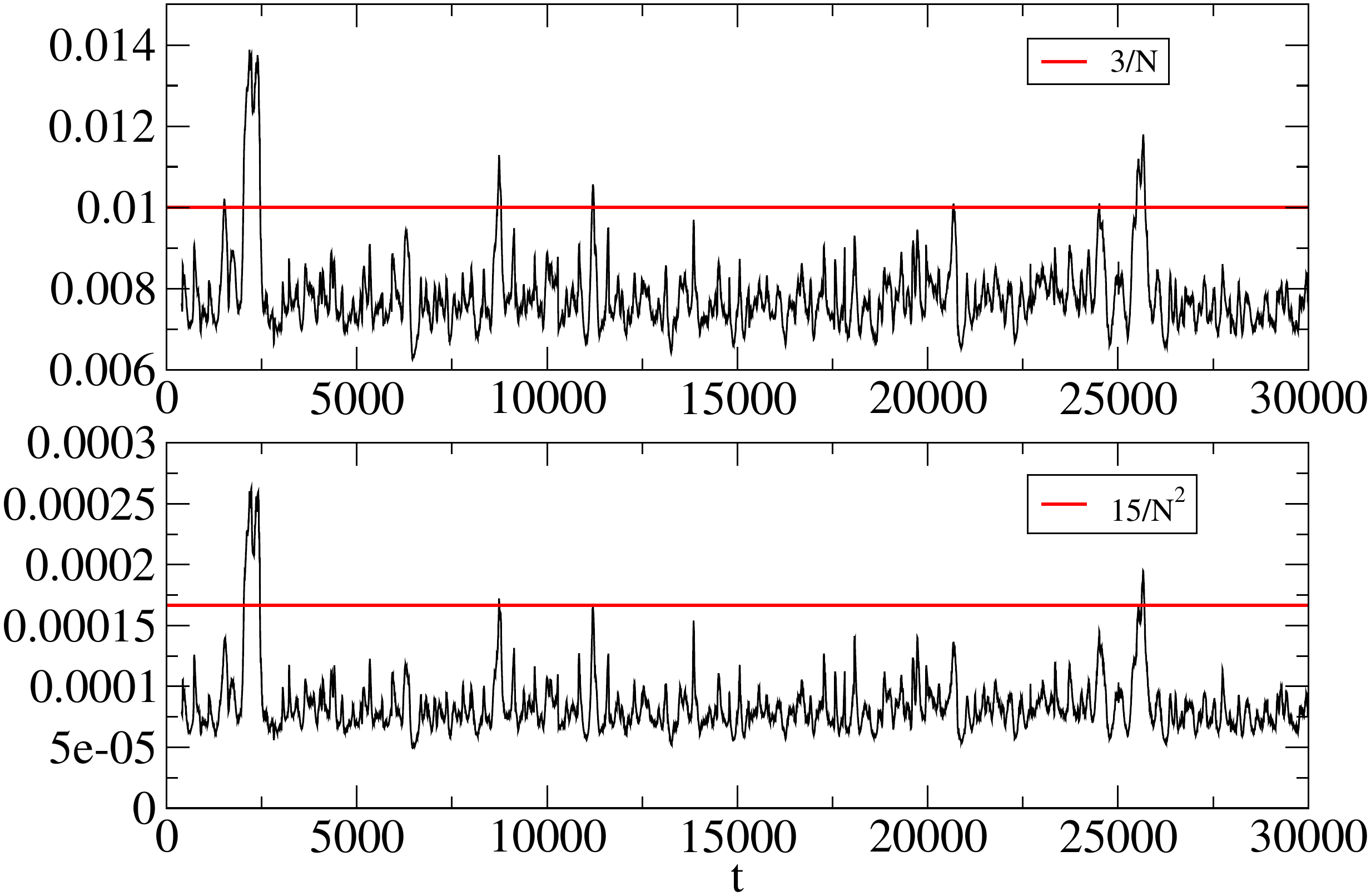}
\caption{
$IPR(1)$ and $IPR6(1)$ from absolute-return eigenvectors.
}
\label{fig:History}
\end{figure}

\begin{figure}
\centering
\includegraphics[height=9.5cm,width=13.cm]{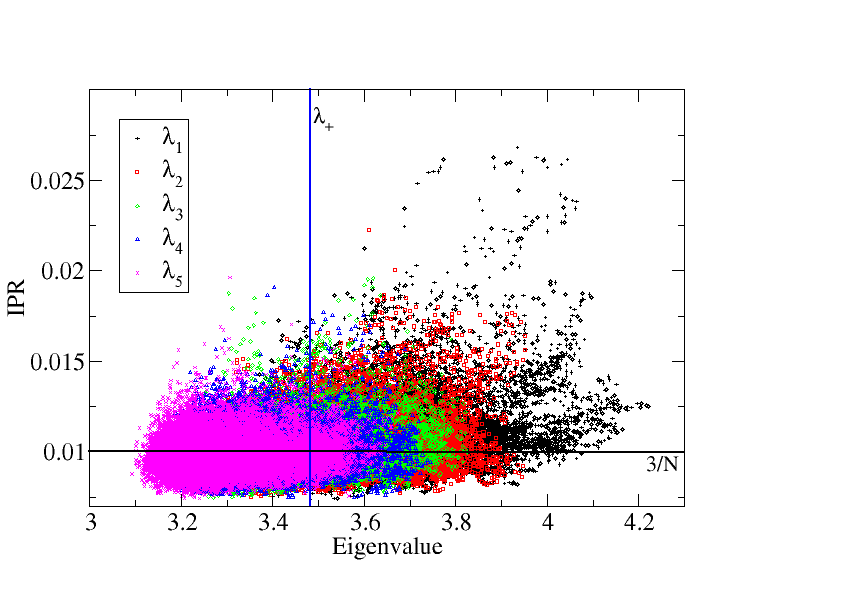}
\caption{
$IPR(l)$ versus return eigenvalues $\lambda_l$, $l=1,\dots,5$. 
}
\label{fig:History}
\end{figure}

\begin{figure}
\centering
\includegraphics[height=9.5cm,width=13.cm]{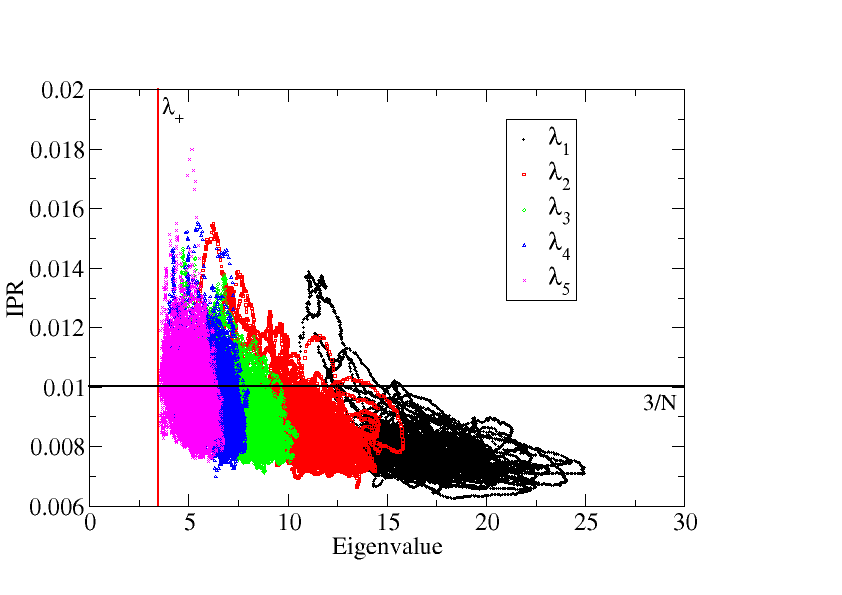}
\caption{
$IPR(l)$ versus absolute-return eigenvalues $\lambda_l$, $l=1,\dots,5$.
}
\label{fig:History}
\end{figure}

Figure 6 shows the time evolution of $IPR(1)$ and $IPR6(1)$ for the return cross-correlation matrix.
$IPR(1)$ and $IPR6(1)$ vary around the values expected from RMT, i.e., $3/N$ and $15/N^2$ respectively,
and no clear correspondence between the volatility index and the IPRs is observed.

On the other hand, $IPR(1)$ and $IPR6(1)$ from the absolute-return cross-correlation matrix in Fig. 7
exhibit different behaviors from those observed in the return cross-correlation matrix.
Namely, the IPRs vary below the expected values from RMT when the volatility index is low,
and the IPRs increase when the volatility index is high.  

Figure 8 shows the scatter plot of $IPR(l)$ versus eigenvalue $\lambda_l$ for $l=1,\dots,5$ for the return cross-correlation matrix.
It seems that the return cross-correlation matrix is already close to the random matrix in this model,
and thus, it might be understood that the IPRs do not clearly show a correspondence with the volatility index.

Figure 9 shows the scatter plot of $IPR(l)$ versus eigenvalue  $\lambda_l$ for $l=1,\dots,5$ for the absolute-return cross-correlation matrix.
For the largest eigenvalue $\lambda_1$, $IPR(1)$ deviates below $3/N$, which is the value expected from RMT.
Moreover, the value of $\lambda_1$ is far beyond $\lambda_+$ from RMT.
Furthermore, $IPR(l)$ gradually approaches the expectation from RMT
for the smaller eigenvalues $\lambda_l, l=2,\dots,5$.

\section{Conclusion}
We performed a large-scale simulation of an artificial financial market
on a $100\times100$ lattice  with 300 stocks using 
the extended Ising-based model.
The financial system simulated by the model exhibited a fat-tailed return distribution and 
volatility clustering similar to real financial markets.
To estimate the volatility level or stability of the financial system,
the volatility index was defined by the average of the absolute-return. This measure showed that the financial system in the simulation underwent several unstable periods.
We also calculated the CRF introduced as a risk measure \cite{SR1}.
We showed that the CRF from the return cross-correlation matrix increases at high volatility periods.
On the other hand, the CRF from the absolute-return cross-correlation matrix decreases at high volatility periods.
It is advantageous to measure the magnitude of changes in the CRF  to identify 
unstable periods rather than to measure the actual changes \cite{SR3}.

RMT was used to estimate the stability of the financial system.
We calculated the IPR and its higher-power version, IPR6, and 
compared them with the results from RMT.
We showed that the IPR and IPR6 from the return cross-correlation matrix were not sensitive to the instability of the financial system.
On the other hand, the IPR and IPR6 from the absolute-return cross-correlation matrix 
increased at high volatility periods.

While the extended Ising-based model exhibits major stylized facts such as a fat-tailed return distribution and
volatility clustering, there exist some inconsistencies with real financial markets.
For instance, in the real financial market,
the IPR from the return cross-correlation matrix changes considerably at unstable periods. 
In addition, the IPR shows a decrease rather than an increase when the market is unstable \cite{SR5}.
Therefore, the model introduced in this study remains a prototype and further studies are needed 
to build  a satisfactory model for real financial markets.

\section*{Acknowledgement}
Numerical calculations in this work were carried out at the
Yukawa Institute Computer Facility and at the facilities of the Institute of Statistical Mathematics.

\end{document}